# Enhanced pinning in high-temperature superconducting cuprate single crystals at low DC magnetic field

*V.Yu.Monarkha, Yu.A.Savina, V.P.Timofeev*

B.Verkin Institute for Low Temperature Physics & Engineering National Academy of Sciences of Ukraine, 47 Lenin Ave., 61103 Kharkov, Ukraine

E-mail: *v.monarkha@ilt.kharkov.ua* ; *timofeev@ilt.kharkov.ua*

**Abstract**

Low density flux dynamics in Nd based cuprate ($NdBa_2Cu_3O_{7-x}$) HTSC single crystals trapped in ≈ 1 Oe range DC magnetic fields were investigated for the first time. Complex measurements of time, temperature and field dependence of magnetization were performed. Using the results obtained and the Anderson-Kim model the normalized magnetization relaxation rate $S$ and the averaged effective pinning potential $U$ were calculated. We show a significant (one or even two orders) increase in the effective pinning potential in these samples in comparison with measurements performed in high magnetic fields for similar cuprate single crystals formerly. In a wide enough range of temperatures close to $T_c$ ($0.72 \leq T/T_c \leq 0.93$) it has been shown that the doping of $NdBa_2Cu_3O_{7-x}$ single crystals with calcium atoms leads to the increase of the averaged effective pinning potential with an insignificant decreasing of the critical temperature.

## 1. Introduction

The high-temperature superconductive (HTSC) single crystals have proven to be a universal object of investigation while studying the magnetic and electrical properties of the solid matter in superconducting state. Rare-earth element based cuprate HTSCs are a wide class of superconductors that are widely used both in fundamental researches of condensed state and in practical applications [1,2]. In spite of an extensive amount of publications on studying the magnetic and electric properties of HTSCs with different crystalline structure the low field and close to critical temperature area of the *H-T* phase diagram remains unstudied [3]. It can be explained by the increase of the thermally activated fluctuation processes near the phase transition domain and high sensitivity sensor required at low DC magnetic fields.

HTSCs are known to be type-II superconductors. The Abrikosov and Josephson vortices and its bundles fix selves on various structural defects with locally suppressed superconducting order parameter. This process is referred to as pinning and $U$ is the effective averaged pinning potential. While the vortexes are pinned the superconductor remains in the dessipationless mixed state. Transport or screening currents affect the vortices in a way similar to the Lorenz force, attempting to drive them off the pinning centers. This effect is enhanced by thermal fluctuations that sway the vortices. The effective pinning potential is starting to decrease. It can be observed as flux creep; dissipation processes and sample magnetization (*M*) decay [4].

The Anderson-Kim thermally activated flux creep model assumes a linear relation of the current density *J* with the creep activation energy, equal to the averaged effective pinning potential depth: $U = U_0(1- J/J_c)$, where $U_0$ is the effective pinning potential in the absence of current, and $J_c$ is the critical (maximal possible dissipativeless) current density. However, the experiments have shown that this dependence is significantly nonlinear and can not adequately describe the superconductor behavior in all temperature ($0 < T < T_c$), magnetic field ($0 < H < H_{c2}$) and current ($0 \leq J \leq J_c$) ranges.

Twinning boundaries (TBs) exist as natural structural defects in cuprate HTSC. They appear in the process of sample fabrication and can be found in single crystals, polycrysatal and thin film samples. The substrate homogeneity additionally strong affects stoichiometry of the thin film samples [5]. The twinning boundaries in HTCS can act as pinning centers at high temperatures and as well as flux creep assisting channels at low temperatures [6]. There is a significant amount of theoretical and experimental publications on the influence of the twinning boundaries on the electrical and magnetic properties of HTSC. However the experimental studies of the pinning effects and magnetic flux dynamics are usually conducted in strong magnetic fields ($\geq 1$ T), when the vortex grid is fully formed and the vortexes strongly interact with each other (vortex matter).

boundaries along $c$ axis through the bulk of the sample and with minimal mosaicity.

Magnetization temperature dependence and isothermal magnetization relaxation curves were obtained using helium-cooled SQUID-gradientometer. We used a standard method of magnetization measurement by registrating the response of a radio frequency SQUID on the sample magnetic moment change in a homogeneous magnetic field of the solenoid. The sensitivity of the experimental set to the minimal magnetic moment of the sample is $\approx 8.6 \cdot 10^{-11} A \cdot m^2$. The effective shielding system reduces the value of external Earth magnetic field inside the measuring chamber to ~ 0.5 mOe. The deviation of the temperature in the chamber from the selected value is ~ 2mK in the 20 -120 K range.

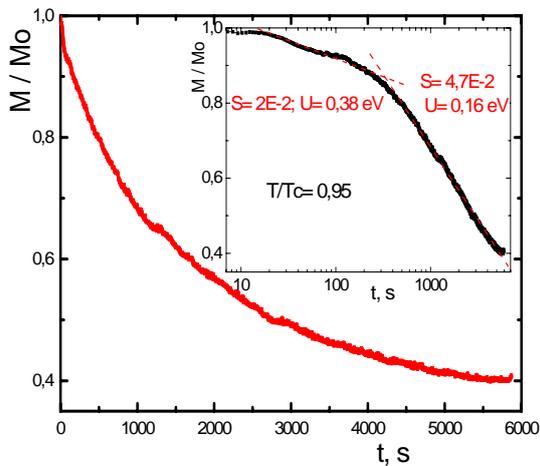

Fig.1. Magnetic relaxation $M$ for one of the tested samples normalized to its initial value $M_0$. Methods for calculation of normalized relaxation rate $S$ and effective pinning potential $U$, respectively according to the linear model of creep for two time windows are shown at inset.

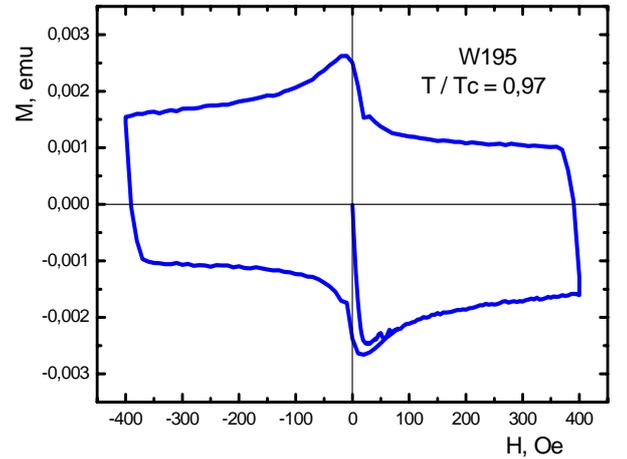

Fig.2. HTSC sample W195 magnetization loop $M(H)$ for the high temperature range.

## 2. Experimental

As the main object of investigation we have chosen cuprate HTSC single crystal samples based on the rare-earth element neodymium ($NdBa_2Cu_3O_{7-x}$), undoped and calcium doped. The typical dimensions of the crystals under study were 0.5-2 mm for $a$ and $b$ axes and 0.015 – 0.1 mm for $c$ axis. The structure of the samples was analyzed using an optical microscope with moving polarized light source. This method allows to clearly observe the unidirectional TB areas and their boarders as well as the mosaic structure of the crystal. To study the influence of the planar defects on the pinning processes we usually selected samples that contained unidirectional twinning

In order to investigate the superconducting phase transition of the sample it was cooled down to ≈ 20 K in the homogenous magnetic field (field cooling, FC-method). The magnetic field was then turned off and the sample was slowly heated up with a constant rate of 0.2 K/min while the magnetization of the sample was recorded constantly. Superconducting phase transition curves obtained, $M(T)$, were used to define the critical temperature $T_c = T_c^{onset}$ of the HTSC crystal under test. While studying the isothermal magnetic relaxation we used the following measurement sequence: the sample was cooled below $T_c^{onset}$ to a fixed temperature in constant homogenous magnetic field. After that the field was turned off. Remanent (due to trapped flux) magnetization change, $M(t)$, was recorded. The curves obtained were used to determine the isothermal relaxation rate $S$ and the effective averaged pinning potential $U$ using various models [4]. Fig.1 shows the typical isothermal relaxation curve for one of the samples and the $S$ and $U$ estimation method:

$$S = 1/M_0 (dM/d\ln t) \text{ and } U = kT/S, \qquad (1)$$

where $M_0$ is the initial magnetization and $k$ is the Boltzmann constant.

To compare different methods of the magnetic flux dynamics investigations in HTSC, we also performed additional measurements of the magnetic properties of cuprate single crystals using a commercial SQUID-magnetometer (Quantum Design, MPMS 5). During those measurements magnetization loops $M(H)$ of the HTSC samples were obtained (Fig.2), as well as $M(T)$ and $M(t)$ characteristics. The magnetization loops allowed us to determine the penetration field $H_p$ and the first (low) critical field $H_{c1}$ of the samples which are difficult to calculate taking into account the geometrical parameters of the sample and boarder effects.

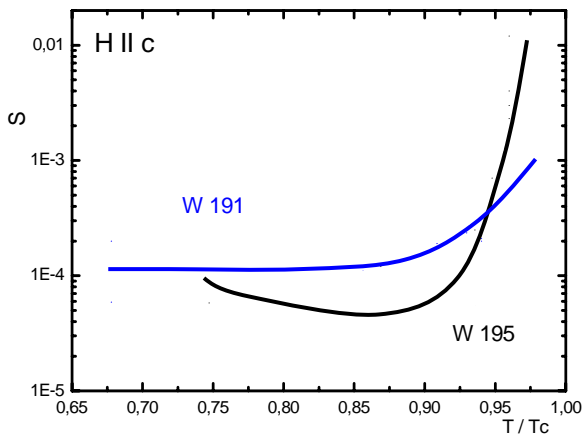

Fig.3. Averaged values of the normalized isothermal relaxation rates $S$ at wide temperature range for two tested samples.

The magnetization curves of the type-II superconductors containing various pinning centers become irreversible and hysteretic. For the typical HTSC samples the presence of pinning centers leads to the transformation of the $M(H)$ curve to a magnetization loop. The width of the magnetization loop $\Delta M$ is proportional to the averaged effective pinning potential. According to Bean's critical state model the critical current density $J_c$ is associated with the geometrical parameters and the $\Delta M$ of the sample under study. The value of $J_c$ can be estimated for example by a relation: $J_c = 15 \Delta M/R$, where $R$ – is the function of the sample geometry which include the demagnetizing factor and the $J_c(H)$ dependence [7].

The experimentally obtained magnetization loops of the single crystals under test are strongly asymmetrical in respect to the $H$ axis which indicates a strong influence of the surface barriers on the vortex dynamics. In these conditions it seams not possible to apply the Bean's critical state model to estimate the critical current density $J_c$ in the selected field and temperature ranges. Thus to estimate the effective pinning potential in low field domain which correspond to the initial part of the magnetization curve $M(H)$, it seams more reasonable to register the magnetization relaxation curves $M(t)$ and use the relation (1) for $U$ value calculation.

## 3. Results and discussion

Fig.3 shows the averaged values of the isothermal relaxation rate for two typical single crystals in a wide temperature range. The sample W191 is undoped ($NdBa_2Cu_3O_{6.98}$), $T_c = 95.3$ K. The sample W195 is doped with calcium ($Nd_{0.88}Ca_{0.12}Ba_2Cu_3O_{6.93}$), $T_c = 82.3$K. The cuprate single crystal doping is used to create artificial point defects which allow to increase the current carrying abilities of the superconductor, $J_c$ [8]. The sample-field orientation was $H \parallel c$ at which the most effective vortex pinning by TBs was observed. It can be seen that the doppant plays the role of additional dot-like pinning centers causing the relaxation rate to drop at a wide temperature range.

The comparison of our results on the relaxation rate behavior with the results obtained for similar HTSCs in significantly higher magnetic fields by other authors [9, 10, and 11] is presented on the Fig.4. The high magnetic fields suppress the random Josephson weak link system in the twinning and grain boundaries thus significantly decrease activation energy for magnetic flux creep and increase the apparent *S* value. Earlier in the experiments with the HTSC samples the growth of the normalized relaxation rate (approximately twice) with the field increasing from ≈ 0 to 1 kOe was observed [12]. Effects associated with the flux pinning in superconductors, have been actively investigated experimentally [13].

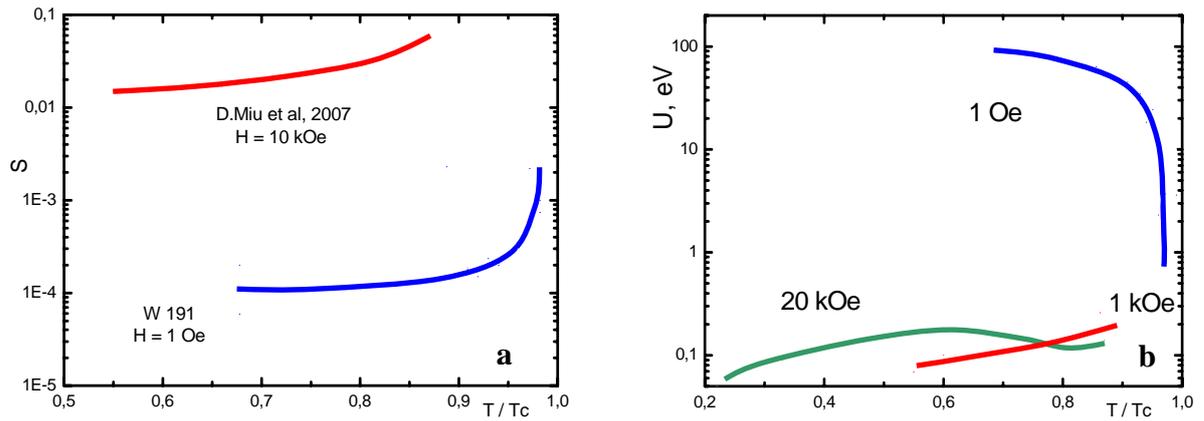

Fig.4. Isothermal magnetization rate *S* drop at the low fields due to TBs effective pinning potential increase in comparison with data from the publication [9] (a) and creep activation energy *U* increase at the low field in comparison with data from publications [10, 11] (b).

## *4. Conclusions*

Low density flux dynamics in Nd based cuprate ($NdBa_2Cu_3O_{7-x}$) HTSC single crystals trapped in 1 - 8 Oe range DC magnetic fields were investigated for the first time. Complex measurements of time, temperature and field dependence of magnetization were performed to achieve this aim. Using the results obtained and relation (1) we were able to calculate the normalized magnetization relaxation rate *S* and the averaged effective pinning potential *U*. We have shown a significant (one or even two orders) increase in the effective pinning potential in these samples in comparison with measurements performed in high magnetic fields for similar cuprate single crystals formerly. It can be explained, for example, by the existence of unperturbed system of randomly distributed Josephson weak links that are present in the volume of the HTSC in twin boundary areas, interblock areas and other planar defects.

In a wide enough range of temperatures close to $T_c$ ($0.72 \leq T/T_c \leq 0.93$) it has been shown that the doping of $NdBa_2Cu_3O_{7-x}$ single crystals with calcium atoms leads to the increase of the averaged effective pinning potential with an insignificant decreasing of the critical temperature.

The results obtained are useful for understanding the nature of pinning and magnetic flux dynamics in layered HTSC cuprates as well as for a practical application in low current superconductive electronics.